\documentclass[twocolumn,9pt]{extarticle}
\RequirePackage{lmodern}
\usepackage[T1]{fontenc}
\usepackage{xcolor}
\usepackage{graphicx}

\usepackage{multirow}

\usepackage{enumitem}
\usepackage{amsfonts}

\usepackage{dcolumn}
\usepackage{bm}
\usepackage[mathlines]{lineno}
\usepackage[lmargin=48.5pt, rmargin=53.5pt, tmargin=40pt, bmargin=60pt, columnsep=20pt]{geometry}
\usepackage{caption}
\usepackage{titling}
\usepackage[superscript, biblabel]{cite}

\usepackage{floatrow}
\usepackage[normalem]{ulem}

\usepackage{physics}
\usepackage{setspace}
\usepackage{comment}
\usepackage[euler]{textgreek}

\pretitle{\begin{flushleft}}
\posttitle{\end{flushleft}}
\preauthor{\begin{flushleft}}
\postauthor{\end{flushleft}}

\makeatletter

\begin{document}

\title{\textsf{\textbf{{\fontsize{25}{60}\selectfont Optical parametric oscillation in silicon carbide nanophotonics}}}}


\author{\large\textbf{\textsf{Melissa A.~Guidry,$^{\dagger}$   Ki Youl Yang,$^{\dagger}$ Daniil M.~Lukin,$^{\dagger}$ Ashot Markosyan, Joshua Yang, Martin M.~Fejer, and Jelena Vu{\v c}kovi\'c}
}}

\date{\small
E.~L.~Ginzton Laboratory, Stanford University, Stanford, California 94305, USA\\
\vspace*{1em}${}^\dagger$These authors contributed equally to this work.
}

\begingroup
\let\center\flushleft
\let\endcenter\endflushleft
\maketitle
\endgroup

\noindent\textsf{\textbf{Silicon carbide (SiC) is rapidly emerging as a leading platform for the implementation of nonlinear and quantum photonics. Here, we find that commercial SiC, which hosts a variety of spin qubits, possesses low optical absorption that can enable SiC integrated photonics with quality factors exceeding $10^7$. We fabricate microring resonators with quality factors as high as $1.1$ million, and observe low-threshold ($8.5 \pm 0.5$~mW) optical parametric oscillation as well as optical microcombs spanning $200$~nm. Our demonstration is an essential milestone in the development of photonic devices that harness the unique optical properties of SiC, paving the way toward the monolithic integration of nonlinear photonics with spin-based quantum technologies.
}}

\vspace*{1em}

\begin{figure*}[ht]
\centering
\includegraphics[width=14cm]{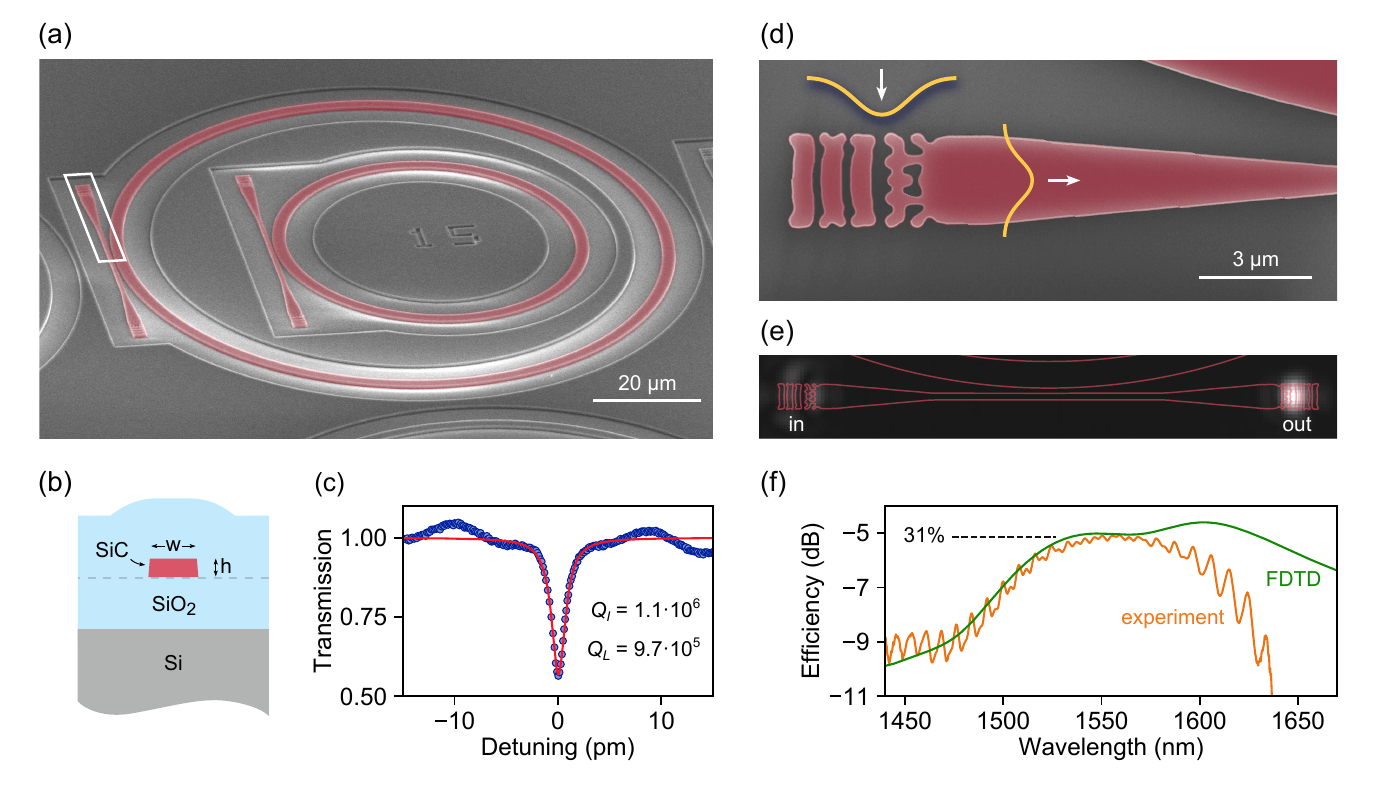}
\caption{\textbf{Fig.~1.} Microring resonators and inverse-designed vertical couplers in 4H-SiC-on-insulator. (a) A scanning electron micrograph (SEM) of two SiC microring resonators (false-colored) with diameters of 55~\textmu m and 100~\textmu m before SiO\textsubscript{2} encapsulation. (b) A schematic of the device cross-section after SiO\textsubscript{2} encapsulation. (c) Transmission spectrum of a ring with diameter 100~\textmu m, width 3.0~\textmu m, and height 530~nm, around a TE\textsubscript{00} resonance  with an intrinsic Q of $1.1\cdot10^6$ and loaded Q of $9.7\cdot10^5$. The wavelength is relative to 1532~nm. (d) A close-up SEM image of the inverse-designed vertical coupler, highlighted in (a). The coupler converts a near-diffraction-limited free-space Gaussian beam (focused via a 50x objective with $\text{NA}=0.5$) into the fundamental waveguide mode. (e) Camera image of the coupler operating at peak efficiency, showing little back-reflection from the input coupler, and a nearly-Gaussian beam at the output. (f) We measure the single-mode coupling efficiency to be 31\% at the target wavelength of 1550~nm, in close agreement with finite-difference time domain (FDTD) simulation.}
\label{fig:1}
\end{figure*}

Silicon carbide (SiC) is a promising material for realizing quantum and nonlinear photonics technologies \cite{awschalom2018quantum, anderson2019electrical, song2019ultrahigh, Lukin2019}. Uniquely combining a wide transparency window (UV to mid-IR) \cite{wang20134h}, a strong second- and third-order ($\chi^{(2)}$, $\chi^{(3)})$ optical nonlinearity \cite{sato2009accurate, Zheng2019}, and a high refractive index, SiC is also host to a variety of optically-addressable spin qubits \cite{Nagy2019Carbide, anderson2019electrical}, which are actively studied for applications in quantum computation  \cite{Nagy2019Carbide,Morioka2020SpinControlled} and sensing \cite{simin2016all}. SiC photonics have been in development for over a decade, and have recently seen major breakthroughs, including low-loss waveguides and high quality (Q) factor resonators \cite{Lukin2019, song2019ultrahigh, fan2020high}; efficient on-chip frequency conversion via the $\chi^{(2)}$ nonlinearity \cite{Lukin2019, song2019ultrahigh}; and integration of single spin qubits with nanophotonic cavities \cite{Lukin2019, crook2020purcell}. This puts SiC on the forefront of efforts toward a monolithic platform combining quantum and nonlinear photonics. However, the observation of optical parametric oscillation (OPO) in SiC remains an outstanding challenge. On-chip OPO enables efficient wideband spectral translation \cite{lu2019efficient}, frequency comb formation for metrology \cite{Newman:2019:Optica} and spectroscopy \cite{Suh:2016:Science}, and on-chip generation of non-classical light states \cite{kues2017chip}.  Furthermore, the monolithic integration of optical spin defects with a near-threshold OPO light source can enable the demonstration of new physical effects in cavity quantum electrodynamics, such as synthetic strong coupling \cite{leroux2018enhancing}, with important implications for integrated spin-based quantum technologies.

Here, we demonstrate on-chip $\chi^{(3)}$ optical parametric oscillation (OPO) and microcomb formation in  high-purity semi-insulating (HPSI) 4H-SiC-on-insulator microring resonators. This is enabled by resonator dispersion engineering, improved fabrication techniques resulting in Q factors as high as 1.1~million, and compact inverse-designed vertical couplers for a broadband, high-efficiency free-space interface. We also perform a careful study of the intrinsic material absorption of SiC, providing crucial information on the dominant sources of loss in high-Q photonic devices based on SiC.

The device fabrication follows the process described in Ref.~\cite{Lukin2019}, with modifications to improve the pattern-transfer fidelity and device Q factors. Instead of using HSQ e-beam resist, which suffers from low reactive-ion etching selectivity against SiC, an aluminum hard mask (deposited via evaporation and patterned with ZEP e-beam resist) is used. Combined with a low-power SF\textsubscript{6} etch, this yields a hard-mask selectivity of 9 (compared to 2 for HSQ). Using this method, devices in SiC films as thick as 1.5~\textmu m can be fabricated. Figure~\ref{fig:1}a shows microring resonator devices before oxide encapsulation. Q factors as high as $1.1\cdot10^6$ are measured (Fig.~\ref{fig:1}c), which corresponds to waveguide loss of 0.38~dB/cm. Routing light to and from the chip is done via efficient and broadband inverse-designed vertical couplers \cite{su2020nanophotonic, dory2019inverse}, with a peak single-mode coupling efficiency of $31\%$, as illustrated in Fig.~\ref{fig:1}(d-f). Accurate pattern transfer and high aspect ratio nanostructures enabled by the new fabrication approach were essential for the demonstration of the close agreement between the simulated and measured efficiency at the target wavelength of 1550~nm.

\begin{figure}[t!]
\centering
\includegraphics[width=8.82cm]{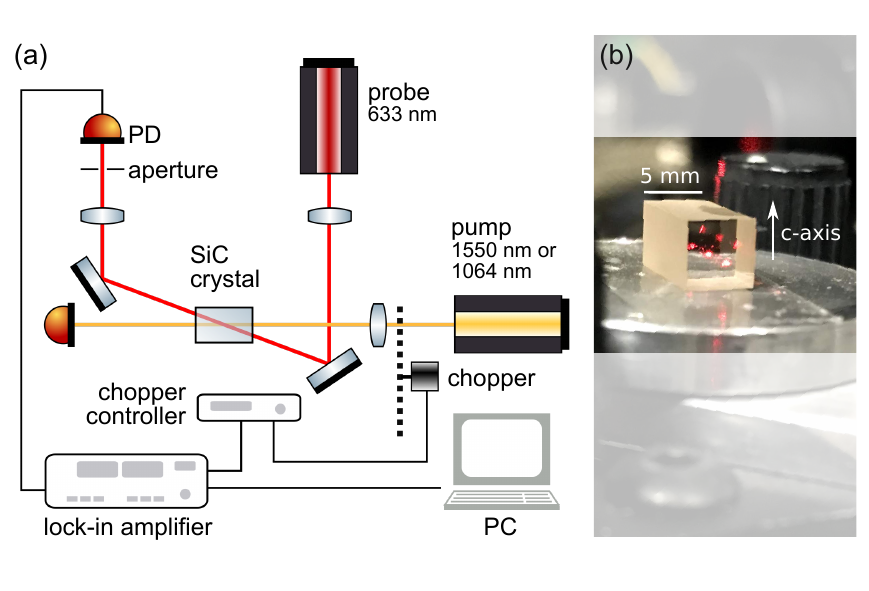}
\caption{\textbf{Fig.~2.} Measurement of the intrinsic loss of 4H-SiC. (a) Diagram of the PCI measurement setup, described in detail in Ref.~\cite{markosyan2013study}. (b) A crystal of 4H-SiC with dimensions of $5\times5\times10$~mm undergoing the absorption measurement. Multiple reflections of the red probe laser inside the crystal are visible.} 
\label{fig:pci}
\end{figure}

The waveguide loss of 0.38~dB/cm presented here approaches the previously reported upper bound of 0.3~dB/cm on the intrinsic absorption of 4H-SiC \cite{cardenas2015optical}. To identify the dominant source of loss in high-Q SiC devices, we perform high resolution characterization of the intrinsic absorption of SiC via photothermal common-path interferometry (PCI), which has been used to detect absolute absorption down to 1~ppm/cm \cite{markosyan2013study}. In PCI, a low-power probe beam is used to sense the heating effect from the absorption of a high-power pump beam, as shown in Fig.~\ref{fig:pci}a. The pump beam, with comparatively smaller waist, is chopped, periodically modulating the heating effect, which induces self-interference of the probe beam via the photothermal effect. We perform absorption measurements on sublimation-grown HPSI 4H-SiC (Shanghai Famous Trade Co. LTD) with resistivity exceeding $10^5$~\textOmega$\cdot$m (Fig.~\ref{fig:pci}b). The measured absorption is shown in Table~\ref{tab:1}. We note that the absolute accuracy of PCI requires a low-transparency calibration sample or precise knowledge of material properties, including the refractive index, the thermo-optic coefficient, the coefficient of thermal expansion, and the thermal conductivity. Based on previously-reported values of these parameters for 4H-SiC \cite{wang20134h, haynes2011crc, wei2013thermal, li1986thermal, watanabe2012thermo}, we conservatively estimate the absolute accuracy of the reported values to be $\pm 25\%$. However, the relative precision is within 1\%. This allows us to observe wavelength-dependent anisotropy in absorption (3.7 at 1550~nm and 2.0 at 1064~nm). Such strong wavelength-dependent anisotropy suggests that residual crystal defects with polarization-dependent near-IR and telecom absorption \cite{zargaleh2018nitrogen, koehl2011room, widmann2015}, rather than the bulk SiC lattice, may be the dominant source of loss; however, further investigation is necessary. Our high-resolution absorption measurements indicate that Q-factors exceeding $10^7$ are possible in SiC. Defect-free epitaxial SiC  layers used in quantum technologies \cite{widmann2015, Nagy2019Carbide, anderson2019electrical} may enable photonics with even higher Q factors.

{\renewcommand{\arraystretch}{1.4}

\begin{table}[t!]
\centering
\caption{{\textbf{Table 1.}} Intrinsic optical loss of HPSI 4H-SiC}
\begin{tabular}{c r|c|cc}
\multicolumn{2}{r|}{\bf Wavelength} & \bf Polarization & \bf Absorption (dB/cm) \\  \cline{2-4} 
&  \multirow{ 2}{*}{ 1064 nm}  &  $\perp$ c-axis & 0.063 & \\
&& $\parallel$ c-axis & 0.031 \\  \cline{2-4}
&\multirow{ 2}{*}{1550 nm} & $\perp$ c-axis & 0.074 \\ 
& & $\parallel$ c-axis & 0.020 
\end{tabular}
  \label{tab:1}
\end{table}

In order to generate degenerate four-wave mixing OPO, one must achieve frequency and phase matching between the pump, signal, and idler modes in the resonator. The frequency matching condition $2 \omega_p = \omega_s + \omega_i$ follows from conservation of energy. The phase matching condition ensures proper volumetric mode overlap and, for OPO within one mode family of a microring, reduces to the statement of conservation of angular momentum $2 \mu_p = \mu_s + \mu_i$, where $\mu$ is the azimuthal mode number \cite{Kippenberg:2004:PRL}. The spectral characteristics of the OPO and subsequent microcomb are determined by the dispersion relative to the pump mode ($\mu_p = 0$)
 \begin{equation}
     \omega(\mu) = \omega_0 + \sum_{k=1} \frac{D_k}{k!} \mu^k
 \end{equation}
 where the $k$\textsuperscript{th}-order dispersion is $D_k$. Here, $D_1$ is the free spectral range (FSR) of the resonator. When $D_2$ dominates all higher-order terms and is positive (negative), the mode dispersion is said to be anomalous (normal). 
 
We engineer microrings to possess anomalous dispersion in the TE\textsubscript{10} mode across the telecommunications band for broadband microcomb generation \cite{Herr2012}. The dispersion calculations include material anisotropy \cite{wang20134h}, and are performed in cylindrical coordinates to include the effect of the microring bending radius. For 100~\textmu m diameter microrings, a target height of 530~nm and a width of 1850~nm (with a sidewall angle of $10^\circ$) are chosen. To predict the OPO behavior, we obtain a transmission spectrum across the full range of the tunable laser (1520-1570~nm), and extract the dispersion of the TE\textsubscript{10} mode by measuring the frequencies of the resonances.  To measure dispersion with high precision, we rely on a Mach Zehnder interferometer ``ruler'', the free spectral range of which is measured using an adaption of the radio-frequency spectroscopy method \cite{Yi:2015:Optica}. Figure~\ref{fig:3}a shows the integrated dispersion $D_{int} = \omega(\mu) - (\omega_0 + D_1 \mu )$  with respect to mode number, to visualize all $k\ge2$ dispersion terms. Numerical simulation of the integrated dispersion for the target microring dimensions is plotted for comparison, showing agreement.

The intrinsic (loaded) Q factor of the TE\textsubscript{10} mode is measured to be $2.7\cdot10^5$ ($1.8\cdot10^5$). At the OPO threshold power, primary sidebands emerge at $\mu = \pm 12$. As more power is injected into the microring, a primary comb at the multi-FSR sideband spacing emerges (Fig.~\ref{fig:3}b). At $75$~mW, spectrally-separated sub-combs are formed around the primary lines. At the maximum injected power, the sub-combs fill out and interfere around the pump, which is evidence of chaotic comb generation \cite{Herr2012}. The thermo-optic effect we observe in our devices may require the use of active capture techniques \cite{Yi:2016:OL} for soliton formation, and lithographic control of device structure can eliminate avoided mode crossings, which may otherwise impede soliton capture. Using the experimental parameters of our device, we simulate the soliton frequency comb using the Lugiato-Lefever equation \cite{Chembo:2013:PRA}, neglecting Raman and $\chi^{(2)}$ effects. The simulated soliton is shown in the last plot in Fig.~\ref{fig:3}b.

\begin{figure}[t!]
\centering
\includegraphics[width=8.82cm]{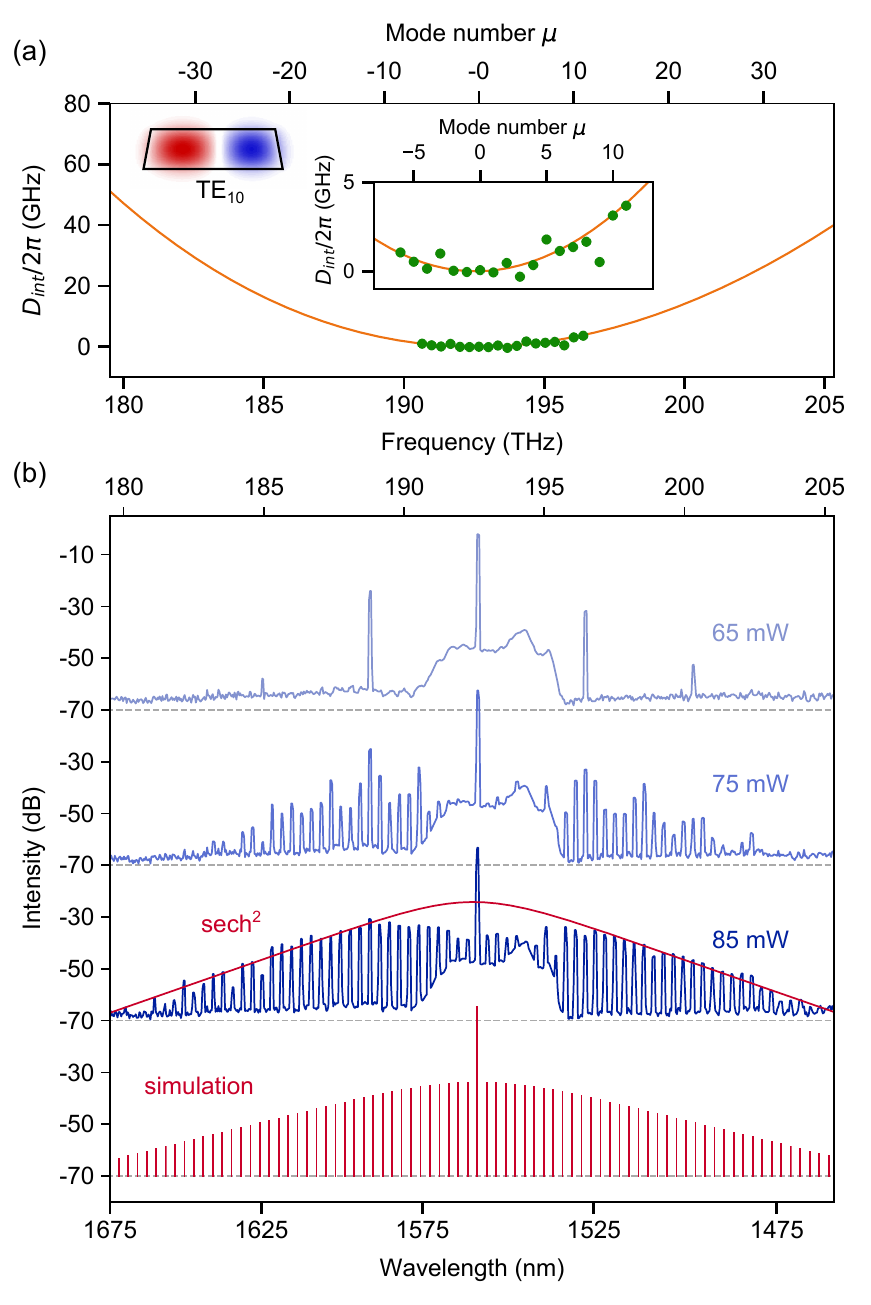}
\caption{\textbf{Fig.~3.} Microcomb formation in a 4H-SiC microring. (a) Measured integrated dispersion (green points) of the TE\textsubscript{10} mode versus the relative mode number $\mu$, where $\mu = 0$ corresponds to the pump mode. The orange curve is a numerical simulation, from which we extract $D_2/2\pi = 61$~MHz and $D_3/2\pi = -0.01$~MHz. Center inset: Close-up of the measured dispersion datapoints. Left inset: Numerical simulation of the TE\textsubscript{10} mode cross-section. (b) Measured OPO spectra (blue) at different injected powers, featuring three distinct stages in the microcomb formation. A sech$^2$ fit (red envelope) is overlaid onto the chaotic frequency comb for comparison to the characteristic soliton spectral shape. Simulation (red) of the soliton frequency comb.}
\label{fig:3}
\end{figure}

Finally, we measure the OPO power threshold in our devices and use it to determine the nonlinear refractive index ($n_2$) of 4H-SiC. The power threshold of the OPO is defined as the power injected into the pump mode at which the primary sideband emerges. This threshold is determined by the loss and the confinement of the three modes
\begin{equation}
    P_{th} = \frac{  \omega_0 n^2 }{8 \eta  n_2 c} \frac{ V }{  \sqrt{Q_{L,s} Q_{L,i}} Q_{L,p}}
\end{equation}
where $n$ is the modal refractive index, $V$ is the mode volume, and $\eta = Q_{L,p}/Q_{c,p}$ where $Q_{c,p}$ accounts for coupling losses from the pump mode to the waveguide \cite{Lu:19}. In this demonstration, we use the TE\textsubscript{00} mode of a 55~\textmu m diameter ring resonator with the same cross-section as before. Although the dispersion is normal for the fundamental TE mode, pumping at an avoided mode crossing allows us to achieve frequency matching \cite{xue2015mode} and to generate OPO, while benefiting from the higher quality factors of the fundamental mode. By optimizing the pump power such that the OPO threshold is reached exactly on resonance, we measure a threshold of $ 8.5\pm 0.5$~mW. Using the simulated mode volume and measured quality factors, we extract a nonlinear refractive index for 4H-SiC of $n_2 = 6.9 \pm 1.1\times 10^{-15} $ cm${}^2$/W at $1550$~nm, consistent with previous studies  \cite{Zheng2019, lu2014optical}. 

In conclusion, we have demonstrated optical parametric oscillation and frequency combs in SiC nanophotonics by leveraging the high field enhancement of our microrings. In light of the recent integration of single spin qubits into 4H-SiC-on-insulator nanostructures \cite{Lukin2019}, our platform holds promise for the monolithic realization of nonlinear and quantum photonics.

\pagebreak

\section*{{\fontsize{11}{11}\textsf{\textbf{Funding Information\vspace{-1.2 ex}}}}}
This research is funded in part by the National Science Foundation under award NSF/EFRI-1741660; and the DARPA PIPES program under contract number HR0011-19-2-0016. Part of this work was performed at the Stanford Nanofabrication Facility (SNF) and the Stanford Nano Shared Facilities (SNSF), supported by the National Science Foundation under award ECCS-1542152. 
M.A.G. acknowledges support from the Albion Hewlett Stanford Graduate Fellowship (SGF) and the NSF Graduate Research Fellowship. K.Y.Y. acknowledges support from a Quantum and Nano Science and Engineering postdoctoral fellowship. D.L. acknowledges support from the Fong SGF and the National Defense Science and Engineering Graduate Fellowship.

\section*{{\fontsize{11}{11}\textsf{\textbf{Acknowledgments\vspace{-1.2 ex}}}}}
We thank Logan Su, Constantin Dory, and Geun Ho Ahn for assistance with inverse design.

 \section*{{\fontsize{11}{11}\textsf{\textbf{Correspondence\vspace{-1.2 ex}}}}} Correspondence and requests for materials
should be addressed to J.V.~(email: jela@stanford.edu).

\section*{{\fontsize{11}{11}\textsf{\textbf{References\vspace{-1.2 ex}}}}}

\begingroup
\renewcommand{\section}[2]{}%
\bibliography{bibliography}

\endgroup


\end{document}